\documentclass[fleqn,usenatbib]{mnras}
\usepackage{newtxtext,newtxmath}
\usepackage[T1]{fontenc}

\usepackage{natbib}  
\usepackage{graphicx}	
\usepackage{amsmath}	

\title[IXPE view of 4U 1630 in the SPL state]{Spectropolarimetric study of 4U~1630$-$47 in Steep Power law state with \textit{IXPE} and \textit{NICER}}

\author[Rawat, Garg $\&$ M\'endez]{Divya Rawat$^{1}$\thanks{E-mail: rawatdivya838@gmail.com (DR)},
Akash Garg$^{1}$,
Mariano M\'endez$^{2}$,
\\
$^{1}$Inter-University Center for Astronomy and Astrophysics, Ganeshkhind, Pune 411007, India\\
$^{2}$Kapteyn Astronomical Institute, University of Groningen, PO BOX 800, Groningen NL-9700 AV, the Netherlands
}

\date{Accepted XXX. Received YYY; in original form ZZZ}

\pubyear{2023}

\begin{document}
\label{firstpage}
\pagerange{\pageref{firstpage}--\pageref{lastpage}}
\maketitle

\begin{abstract}
We probe the spectropolarimetric properties of the black-hole binary source 4U~1630$-$47 in the steep power law state. We detect a significant polarization fraction of $\sim$7 \% at a polarization angle of $\sim$21 $^\circ$. The $2-12$ keV {\it {NICER}} spectrum can be fitted with a combination of a thermal and a Comptonization component, the latter characterized by a spectral index, $\Gamma \sim$2.1, along with a reflection feature at $\sim$7.0 keV. In the $2-8$ keV band, the degree of polarization of 4U~1630$-$47 in the steep power law state is 4.4 $\sigma$ different from the value previously measured in the high soft state. In the steep power law state, the polarization fraction increases as a function of energy but exhibits an overall drop in each energy band compared to that of the high soft state. We propose that the decrease in the polarization fraction in the steep power law state could be attributed to the presence of a corona. The observed polarization properties in both states can be explained by the self-irradiation of the disk around a Kerr black hole, likely influenced by the frame-dragging effect.

\end{abstract}




\section{Introduction}

 Throughout their outburst, black-hole X-ray binary (BXB) systems undergo a transition across various spectral states, commencing with the low-hard state (LHS) and progressing towards the high-soft state (HSS) via the hard and soft intermediate states \citep[][and references within]{be05,be11}. In addition to the spectral states mentioned earlier, at luminosities close to the Eddington limit, certain BXBs exhibit a spectral state called Very high State/Steep power law state \citep[SPL hereafter, see][]{re06,mo12}. While either the disk or the power-law components are responsible for the primary emission in, respectively, the HSS and LHS, the SPL state is distinguished by the coexistence of both thermal and power-law components (\citealt{gi10} and references within). The SPL spectrum can extend upto $\sim$1 MeV and is characterized by a strong power-law component with $\Gamma$ $\ge$ 2.5 \citep{mi91}.  
 
4U~1630$-$472 (hereafter 4U 1630) is a transient low-mass X-ray binary source discovered almost five decades ago with {\it{Uhuru}}, {\it{Ariel V}} \citep{jo76} and {\it{Vela 5B}} X-ray monitor \citep{pr86}, independently.
 This source shows particularly soft outbursts with a lack of hard X-ray emission \citep{ca15} and multiple iron absorption lines in the 6–9 keV band \citep{ku07,ro14,mi15,ra23}. There is a discrepancy in the measurement of the inclination angle of the source; \citet{ku98}, and \citet{ki14} reported it as a high inclination source ($\sim$60--70$^{\circ}$), while \citet{ro14} proposed that it is a low inclination system ($i \sim (11 \pm 5)^\circ$). 
 
 Extensive investigations into the X-ray spectra of 4U 1630 both in the HSS and SPL state have been conducted using various X-ray missions \citep{ku07,ho14,mi15,wa16,pa18}. However, the corona geometry can not be obtained from the fits to the energy spectra. In light of this, the launch of the {\it {IXPE}} mission provides an opportunity to investigate the evolution of spectral transitions through X-ray polarimetry and probe the corona geometry. Previously, \citet{ra23,Kushwaha_2023,Ratheesh_2023} reported a significant polarization of 8\% for 4U 1630 in the HSS with {\it {IXPE}}. Similarly, \citet{po23} and \citet{do23} have observed polarization fraction of $\sim$ 1.1\% for LMC X-1 and  $\sim$ 2.27\% for Cygnus X-1 in the HSS and the variable soft state respectively.\\

  In this work, we conduct the first spectro-polarimetric study of the transient black hole binary source 4U 1630 in the steep power-law state. We describe the observations and data analysis in Section \ref{sec_obs} and the polarimetric and spectro-polarimetric results in Section \ref{sec_res}. In Section \ref{sec_concl}, we discuss our findings, focusing specifically on the comparison of the spectro-polarimetric properties of the source both in the high soft and steep power law states.

\section{Observation and Data Analysis}\label{sec_obs}
\begin{figure}
\centering\includegraphics[scale=0.4,angle=0]{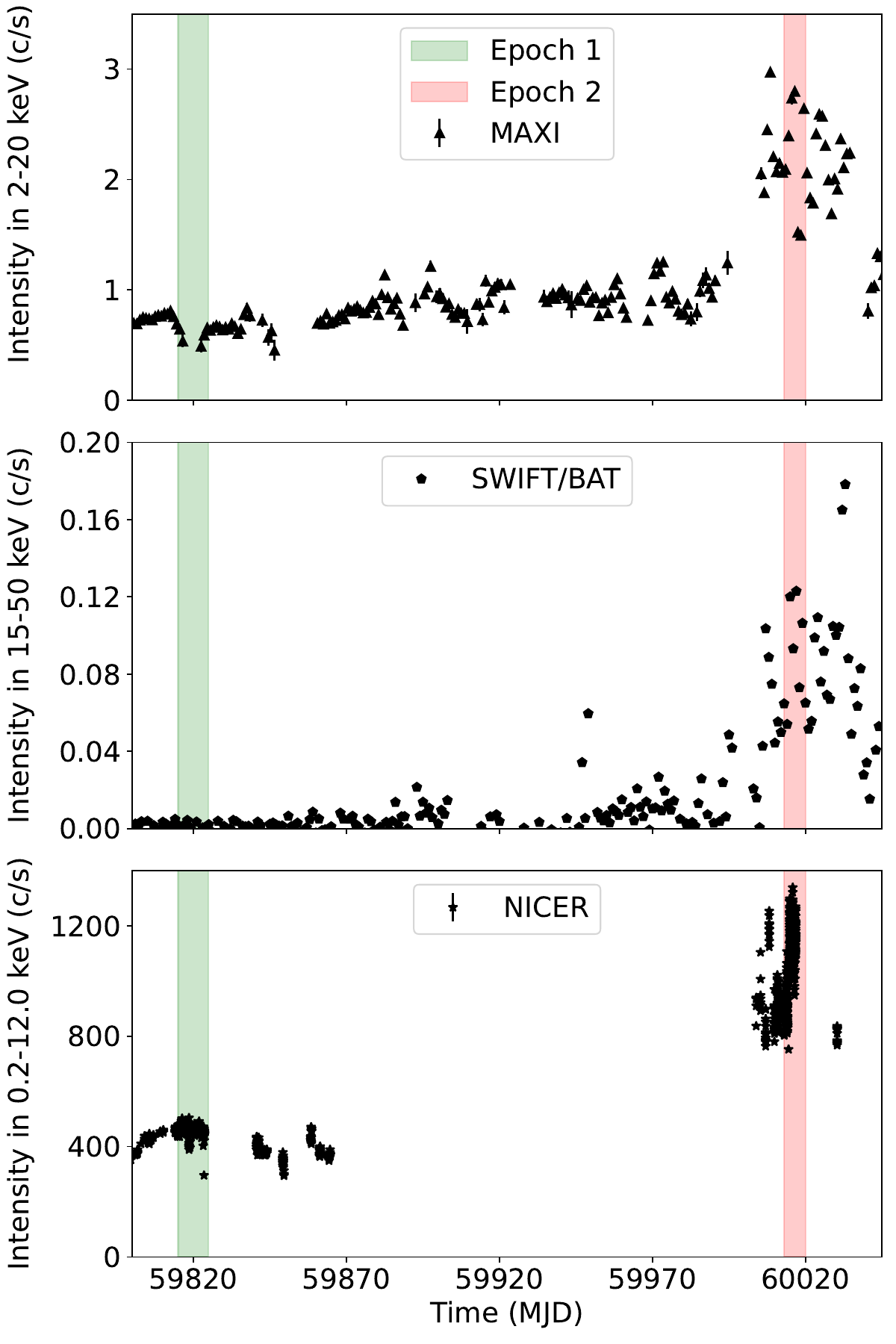}
\caption{{\it{MAXI}}(top panel),  {\it{Swift/BAT}}  (middle panel) and \textit{NICER} (bottom panel) light curves of 4U~1630$-$47 in the $2-20$~keV, $15-50$~keV, and $0.2-12.0$~keV band, respectively. The shaded area represents the duration of the {\it{IXPE}}  observations of the source.}
\label{lightcurve}
\end{figure}

\textit{IXPE} is an X-ray polarimetry mission built under NASA's Astrophysics Small Explorer (SMEX) program in partnership with the Italian Space Agency (ASI). The \textit{IXPE} payload consists of three co-aligned and independent telescopes, each containing a Mirror Module Assembly (MMA) and a Detector Unit (DU). The MMA focuses the X-ray photons onto the DU, which then measures the polarization along with the spatial, temporal and spectral information of those photons. The acquired information is passed to the Detector Service Unit (DSU) to be further transmitted to the ground stations. Each DU has a Gas pixel detector to detect X-ray photons in the range of 2-8 keV and image their resulting short photoelectric tracks  \citep{ma20,ba21,di22,we22}. 
\begin{table}
 \caption{Observation log for epoch 2 of 4U~1630$-$47. The columns are the Instruments used, their ObsID, the observation's start and end date, and the exposure times.}
 \begin{center}
\scalebox{1.0}{%
\begin{tabular}{ccccc}
\hline
Instrument &ObsID & Tstart & Tstop & exposure \\
& & (M.J.D) & (M.J.D) & secs\\ 
\hline

NICER  & 6130010107 & 60013.05 & 60013.06 & 949 \\
& 6557010101 &  60013.76 & 60013.97 & 4819 \\ 
& 6557010102 & 60014.01 & 60014.16 & 3265\\
  & 6557010201 & 60014.47 & 60015.00 & 4927\\
 & 6557010202 & 60015.05 & 60015.72 & 5576\\
 & 6557010301 & 60015.76 & 60015.97 & 5490\\
 & 6557010302 & 60016.02 & 60016.81 & 8862\\
\hline
IXPE & 02250601 & 60013.81 & 60016.79 & 137717 \\
\hline
\end{tabular}}
\end{center}
\label{obs_log}
\end{table}
\textit{IXPE} observed the black hole system 4U 1630 on two instances, 2022 August 23 - 2022 September 02 (Epoch 1) and 2023 March 10-13 (Epoch 2). Table \ref{obs_log} lists the observation IDs for Epoch 2. The top, middle and bottom panels of Figure \ref{lightcurve} show the MAXI (2-20 keV), SWIFT/BAT (15-50 keV) and \textit{NICER} (0.2-12.0 keV) lightcurves, respectively, along with the simultaneous \textit{IXPE} observations in the two Epochs. The $0.2-12.0$ keV \textit{NICER} lightcurve exhibited a fractional rms variability of $\sim$6.5 \%. Clearly, Epoch 1 is a low flux state, whereas Epoch 2 is a high flux state. For Epoch 1, \citet{ra23}, \citet{Kushwaha_2023} and \citet{Ratheesh_2023} independently reported a significant polarization degree of $\sim 8 \%$ at a polarization angle of $\sim 18^\circ$ in the 2-8 keV band. In this work, we investigate the \textit{IXPE} data in conjunction with simultaneous \textit{NICER} data for Epoch 2.

\begin{figure}
\centering\includegraphics[scale=0.38,angle=0]{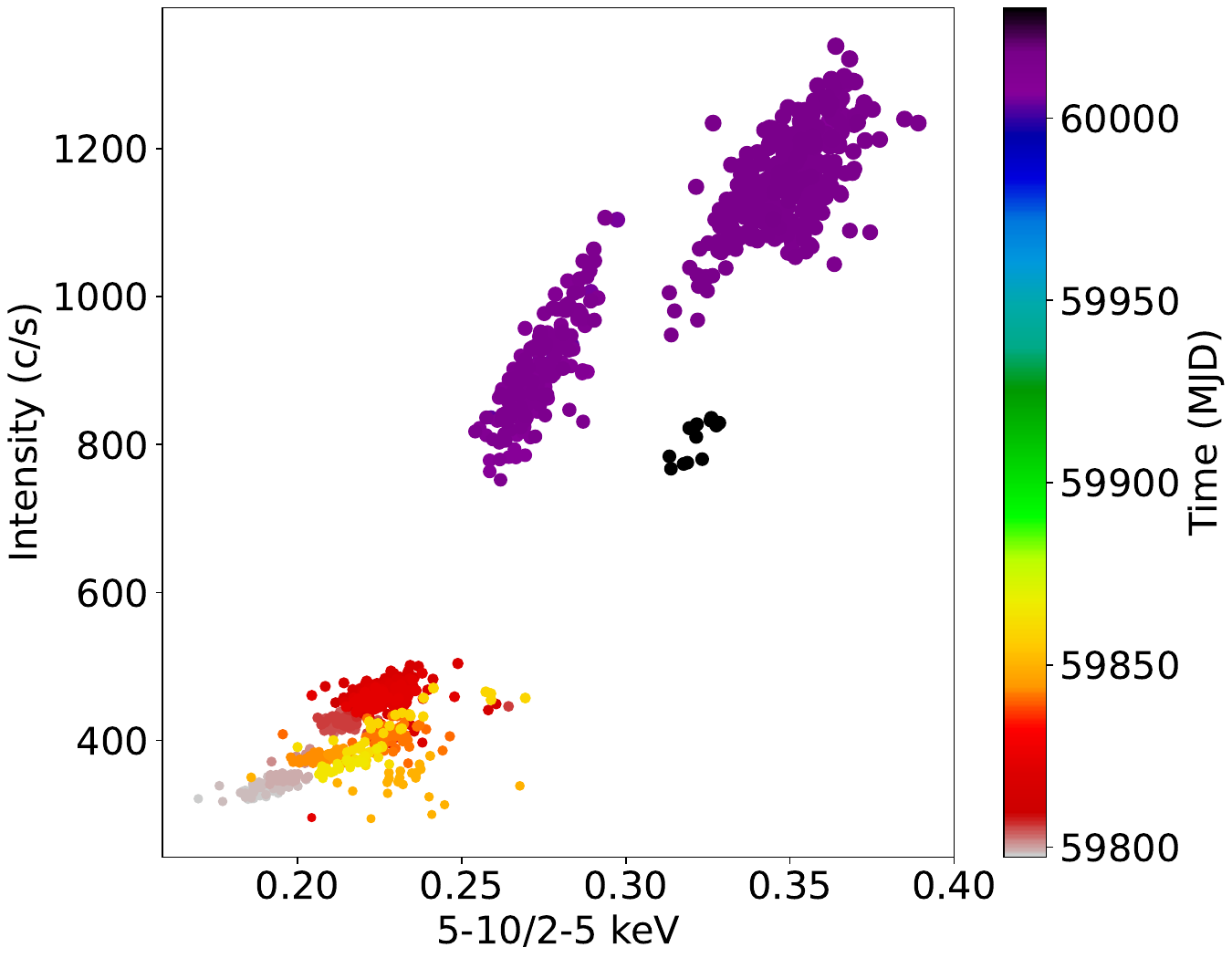}
\caption{The {\textit{NICER}} HID of 4U~1630$-$47. The red points correspond to Epoch 1 in the HSS, while the purple points correspond to Epoch 2 in the SPL state.}
\label{hid}
\end{figure}

We analyzed the \textit{IXPE} data using the Python based software {\sc{ixpeobssim}} 30.2.2\footnote{\label{note2}\url{https://ixpeobssim.readthedocs.io/en/latest/overview.html}} \citep{ba22}. {\sc{ixpeobssim}} provides multiple Python scripts to analyse \textit{IXPE}'s level2 event files. Specifically, we used one of the tools, {\sc{xpselect}}, to generate separate cleaned event files for the source and background for all DUs. We took a circular region with a radius of 60'' for the source and an annular region with an inner radius of 180'', and an outer radius of 240'' for the background. Further, we utilized the tool {\sc{xpbin}} to simplify the event lists by binning with different algorithms \citep{kis15}. Using the algorithm {\sc{pcube}}, we calculated the model-independent polarization degree and polarization angle from the measured azimuthal angle distributions of photo-electrons. We then generated {\sc{xspec}} readable count \textit{I} and Stokes \textit{Q} and \textit{U} spectra for all DUs by applying the algorithms {\sc{PHA1}}, {\sc{PHA1Q}} and {\sc{PHA1U}} respectively. All the {\sc{xpbin}} products are produced in the energy range 2-8 keV using the response functions v012 of {\sc{ixpeobssim}}. We remark here that the background correction was done by scaling down the {\sc{pcube}} parameters to the extraction area and then subtracting them from the parameters of the source.\\
\textit{NICER} monitored the source 4U 1630 frequently since it went into an outburst in 2022. For Epoch 2, we took the \textit{NICER} observations starting from 2023 March 10-13. The details of the observation IDs, start and end time of each observation, with exposure times for the overlapping periods, are given in Table \ref{obs_log}. For each observation, we used the {\sc{nicerl2}} task to apply the standard calibration and screening to the \textit{NICER} level2 data. Subsequently, we utilized the {\sc{nimpumerge}} task to merge the resulting clean event files. Figure \ref{hid} displays
the hardness-Intensity diagram (HID) with the colour bar representing the time in MJD units. The intensity in the HID is defined as the background-subtracted count rate in the $0.2-12$ keV band, and the hardness is the ratio of the background-subtracted count rates in the $5-10$ keV band and the $2-5$ keV bands. During Epoch 2 (around 60,000 MJD), it appears that the source is in a high-flux and spectrally hard state. 

\begin{figure}
\centering\includegraphics[scale=0.38,angle=-90]{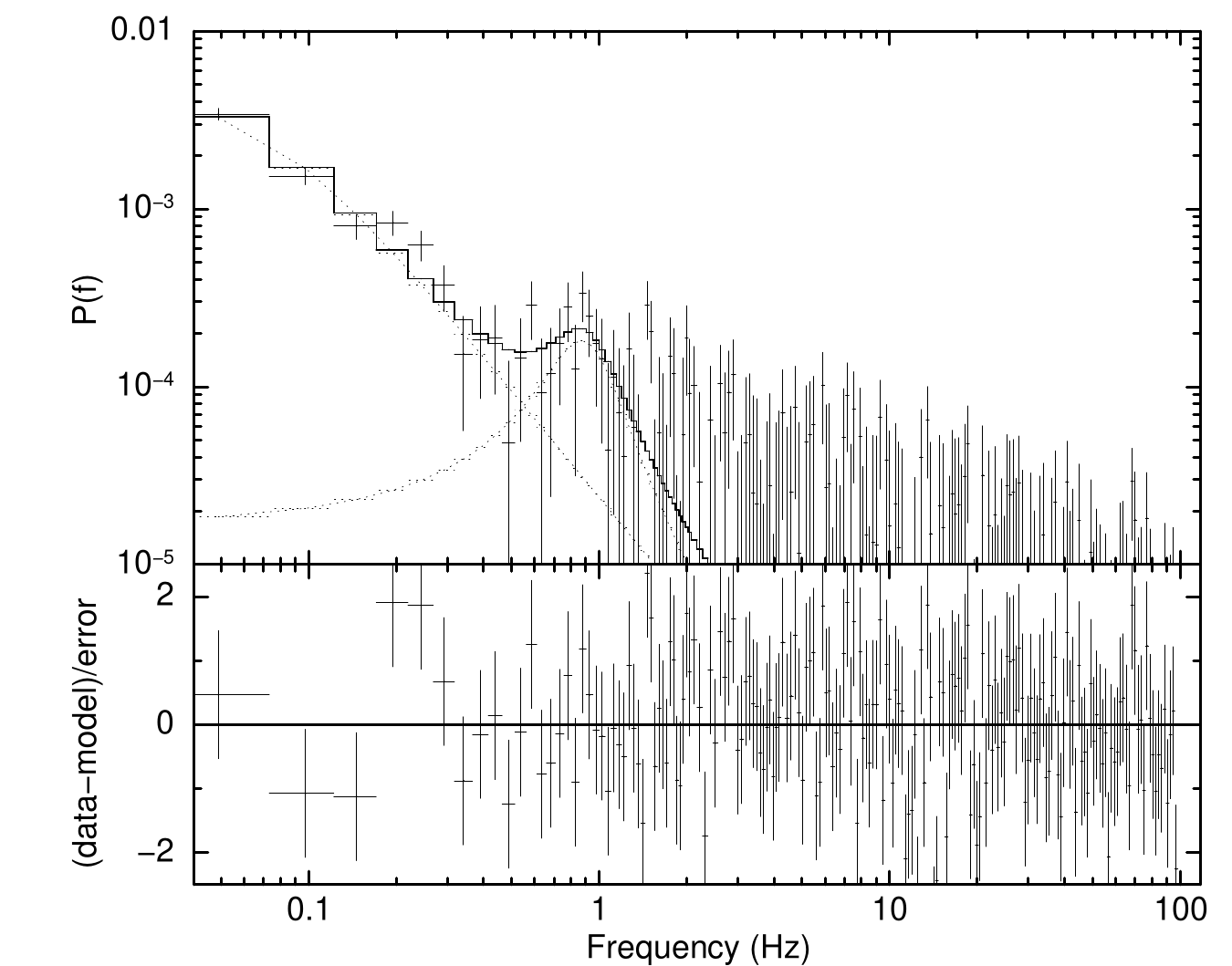}
\caption{The top panel shows PDS of 4U~1630$-$47 fitted with two Lorentzians for {\textit{NICER}} obs ID 6557010302. The bottom panel shows the respective residuals of the fits.}
\label{pds}
\end{figure}

 Using the General High-energy Aperiodic Timing Software (GHATS)\footnote{\url{http://www.brera.inaf.it/utenti/belloni/GHATS_Package/Home.html}}, for each {\it {NICER}} observation we computed Power density spectra (PDS) normalized to units of rms squared per Hz \citep{be90}. We divided the 0.2–12.0 keV lightcurve into 1127 intervals, with each interval of length 21.3 seconds. The time resolution within each interval is set at 2.6$\times$ 10$^{-3}$ s, resulting in a Nyquist frequency of 192.3 Hz. Further, an averaged Poisson noise subtracted power spectrum was produced by averaging all the 1127 PDS and rebinned logarithmically. As shown in Figure \ref{pds}, we fitted the PDS with two Lorentzian functions - one zero-centred Lorentzian with a width of $\sim$0.14 Hz and rms $\sim$2.3 \% and a very broad Lorentzian at a frequency of $\sim$0.9 Hz with a width of $\sim$0.6 Hz and rms $\sim$1.2 \%. Both components represent any possible noise components from the source.

We then extracted the \textit{NICER} spectra and background files using the \textit{NICER} background estimator tool {\sc{3C 50\_RGv7}} \citep{re22}. We used {\sc{Heasoft}} version 6.31.1\footnote{\url{https://heasarc.gsfc.nasa.gov/lheasoft/download.html}} and {\sc{CALDB}} version 20221001\footnote{\url{https://heasarc.gsfc.nasa.gov/docs/heasarc/caldb/nicer/}} to create the response (rmf) and ancillary response (arf) files. To account for instrumental uncertainties, we added a systematic error of 0.8 $\%$. We applied the optimal binning over the \textit{NICER}'s count \textit{I}, \textit{IXPE}'s Stokes \textit{Q} and \textit{U} spectra using {\sc{ftgrouppha}}. Additionally, we grouped the background fits file with Stokes spectra using {\sc{grppha}}. We perform the spectro-polarimetric fitting in the X-ray spectra fitting package {\sc{Xspec Version 12.13.0}} \citep{ar96}.
 \begin{figure}
\centering\includegraphics[scale=0.5,angle=0]{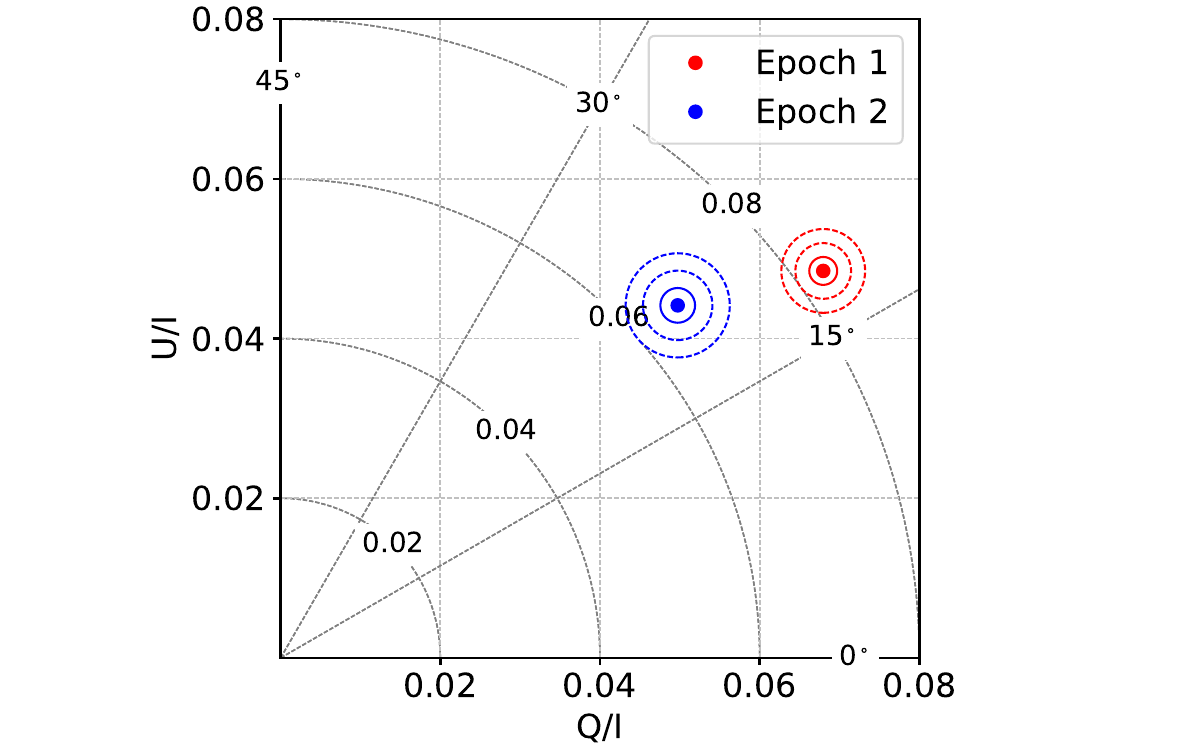}
\caption{Contour plots of PA and PD of 4U~1630$-$47 for Epoch 1 (HSS) and Epoch 2 (SPL state). Contours in the $2-8$ keV bands are derived by summing events from all DUs using the {\sc{pcube}} algorithm, representing confidence levels of 68.27\%, 95.45\%, and 99.73\%.}
\label{2_8_stoke}
\end{figure}
\section{Results}\label{sec_res}
We carried out the polarization analysis of 4U 1630 using two different and independent approaches. The first one is an unweighted and model-independent procedure, {\sc{pcube}}, where the Stokes and polarization parameters are estimated from the emission angles of the photo-electron tracks. The second one is a model-dependent method where a combination of polarization and radiative models are fitted to the count \textit{I}, and Stokes \textit{Q} and \textit{U} spectra in {\sc{xspec}}.

\subsection{Polarization analysis using {\sc{pcube}} method}
\begin{figure}
\centering\includegraphics[scale=0.5,angle=0]{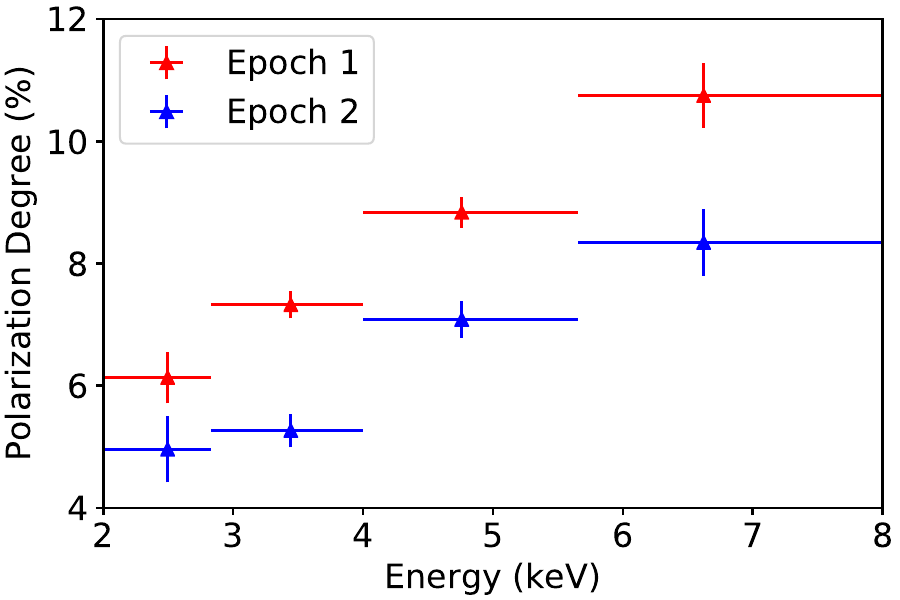}
\caption{The polarization degree as a function of energy for Epoch 1 (HSS) and Epoch 2 (SPL state) of 4U~1630$-$47.}
\label{ene_pol}
\end{figure}

For Epoch 1, \citet{ra23} found a significant energy-dependent polarized emission in 4U 1630, with a polarization degree (PD), $ 8.3 \pm 0.3 \%$ at a polarization angle (PA), $17.8^\circ \pm 1.1^\circ$ in the full energy band of $2-8$ keV during the HSS of the source. For Epoch 2, when the source transited to the SPL, we measured a PD of $6.7\pm 0.2\%$ in the same energy band, different from PD in Epoch 2 at the $4.4\sigma$ level. The measured polarization angle is $20.8^{\circ} \pm 0.9^{\circ} $, consistent with the one in Epoch 1. Figure \ref{2_8_stoke} presents the contour plot of the polarization parameters for both Epochs, confirming the difference in PD between the two Epochs. Table \ref{pol_table} lists the values of the polarization parameters for all DUs for Epoch 2. During the revision of this work, a preprint by \citet{ro23} presents similar values of PA and PD for Epoch 2. While extracting the {\sc{pcube}} parameters in four energy bands, we found a strong energy dependence of the polarization degree in Epoch 2, as shown in Figure \ref{ene_pol}. This is similar to what was found in Epoch 1. The bottom panel of Figure \ref{lightcurve} shows that in Epoch 2, the average {\it{NICER}} count rate changes from  $\sim$930 counts/sec to $\sim$1160 counts/sec. To further explore any flux dependence of the polarization properties, we divided the data of Epoch 2 into two segments based on the {\it{NICER}} flux level, and we found that the polarization properties are consistent with being the same within errors in the two segments.

\subsection{Spectropolarimetric analysis}
\begin{figure}
\centering\includegraphics[width=0.3\textwidth,height=0.35\textheight,angle=-90]{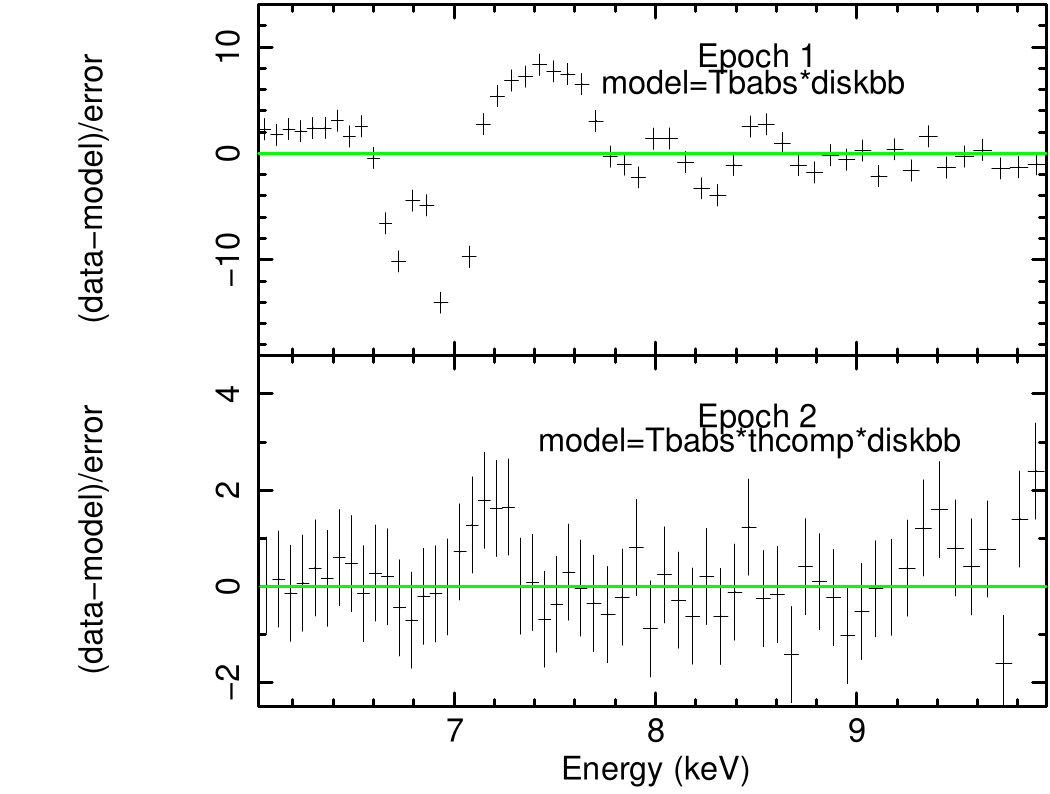}
\caption{The top panel shows residuals of \textit{NICER} spectra of Epoch 1 (HSS) for 4U~1630$-$47 fitted with the model {\sc{tbabs*diskbb}} in the $6-10$ keV band. The bottom panel shows the residuals of the fit of Epoch 2 (SPL state) \textit{NICER} spectra, which were modelled using the model {\sc{tbabs(thcomp*diskbb)}}.}
\label{nicer_spectra}
\end{figure}

In the previous section, using the {\sc{pcube}} method, we found that the polarization properties do not vary within Epoch 2. So, we have merged all the {\it{NICER}} observations listed in Table \ref{obs_log} for the spectropolarimetric analysis.
For Epoch 1, \citet{ra23} modelled the \textit{NICER} spectrum with an absorbed thermal disc component only and detected absorption lines in the $6-9$ keV band, which indicated the presence of disc winds. Figure \ref{nicer_spectra} (top panel) shows the residuals of the \textit{NICER} spectrum for Epoch 1 in $6-10$ keV when fitted with the model {\sc{tbabs*diskbb}}. \citet{ra23} also noted the absence of any significant hard emission, thereby confirming that the source was in the HSS. In this work for Epoch 2, we initially fitted the \textit{NICER} spectrum to obtain a preliminary estimate of the source spectral properties. Since the source is highly absorbed \citep[$N_H \sim$$8-12\sim 10^{22}$ cm$^{-2}$,][]{mi15,wa16,ga19,ra23}, we excluded the data below 2 keV (please see the analysis section of \citealt{ra23}). Interestingly, we found that in this Epoch, the spectrum in the $2-12$ keV band can be described with a non-thermal Comptonized component along with thermal disc emission, revealing that the spectrum of the source was harder than in Epoch 1. For the fit, we used the {\sc{xspec}} model {\sc{tbabs*(thcomp*diskbb)}} with the abundance tables of \citet{wi00} and the cross-section tables of \citet{ve96} for the {\sc tbabs} component. For a systematic error of 0.8 \%, the spectral fitting gave an absorption column density $N_H \sim$11.5 $\times$ $10^{22}$ cm$^{-2}$ and $\chi^{2} = 179.0$ for 159 dof. The residuals at  $\sim$2.4 keV are likely from the reflectivity of the gold M shell edge in the instrument effective area\footnote{\label{nicer_arf}\url{https://heasarc.gsfc.nasa.gov/docs/nicer/analysis_threads/arf-rmf/}}. We also observed a reflection feature at $\sim$ 7.0 keV, as shown in the residuals of the fit in the bottom panel of Figure \ref{nicer_spectra}. We added a {\sc{gaussian}} component to the model to take care of this feature, and the fit gives a $\chi^2$ of  $161.9$ for 156 dof. We still see a residual at $\sim$6.7 keV and added one more {\sc{gaussian}}, the best fit gives a $\chi^2$ of  $157.7$ for 153 dof.

\begin{figure*}
\centering\includegraphics[scale=0.6,angle=-90]{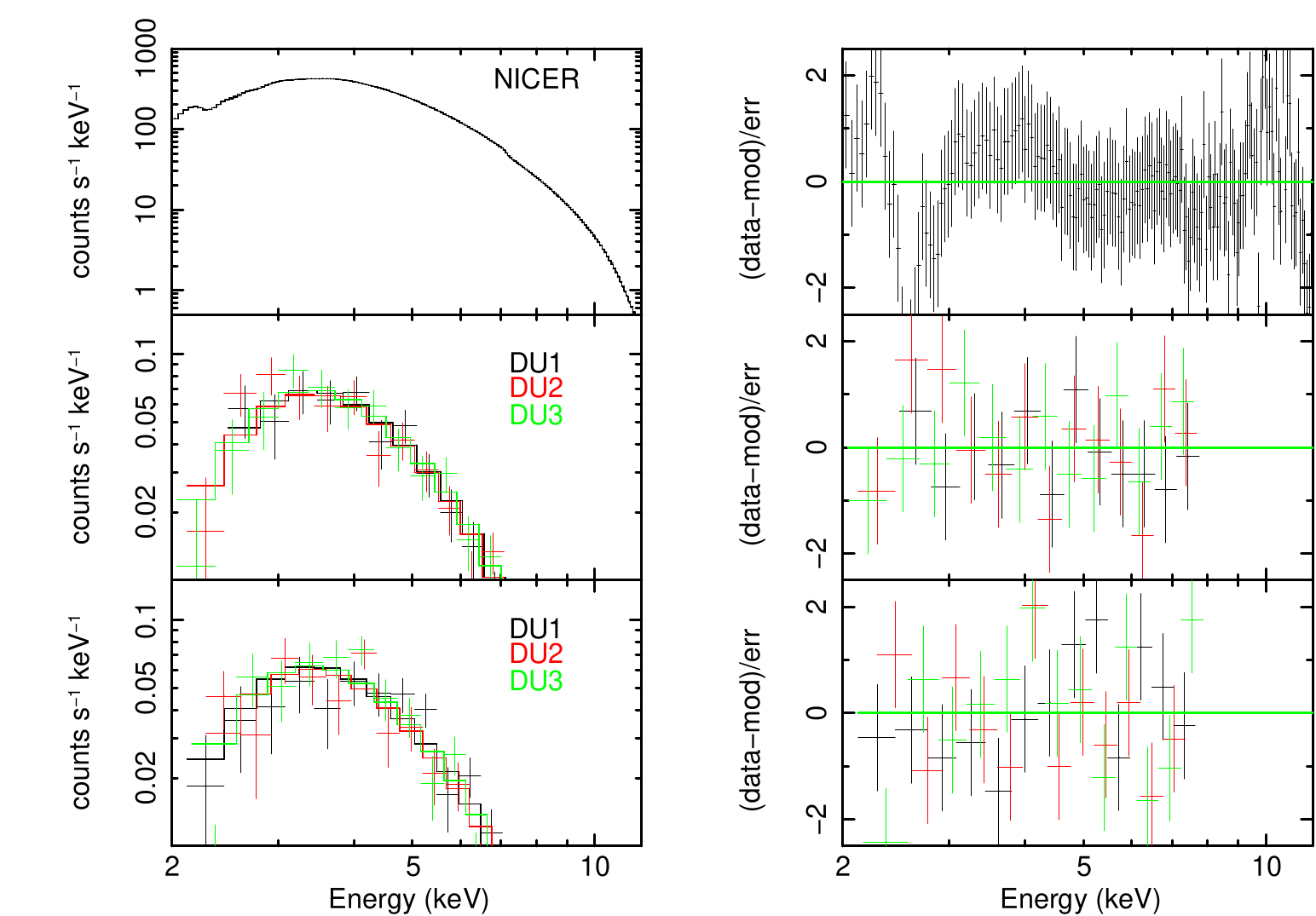}
\caption{The left panels present the results of the simultaneous fits to the {\it{NICER}} spectra (top left panel) and {\it IXPE} Stokes spectra, $Q$ (middle left panel), and $U$ (bottom left panel) of 4U~1630$-$47 in the SPL state. The model used for fitting the data is {\sc{polpowtbabs(thcomp*diskbb+gaussian+gaussian})}. The right panels display the residuals of the best-fitting model. For the best-fitting parameters, refer to Table \ref{spectra_table}.}
\label{spectra_I_Q_U}
\end{figure*}

Further, we fitted the simultaneous {\it{NICER}} spectrum in the $2-12$ keV band and {\it{IXPE}}  $Q$ and $U$ Stokes spectra in the $2-8$ keV band in {\sc{xspec}}. Since the polarization fraction is energy-dependent, we utilized the polarization model {\sc{polpow}} along with the same radiative model combination as taken above for the \textit{NICER} spectrum (see section 3.2 in \citealt{ra23}). We therefore considered the full model combination {\sc{polpow*tbabs*(thcomp*diskbb+gaussian+gaussian})} for the spectro-polarimetric fitting where the component {\sc{polpow}} is $\mathrm{PD}(E)=A_{\rm norm} \times E^{-A_{\rm index}}$ (in fractional units) and $\mathrm{PA}(E)={\psi_{\rm norm}} \times E^{-{\psi}_{\rm index}}$ (in degrees). As the polarization angle is nearly constant with energy at 21$^\circ$, we fixed ${\psi}_{\rm index}$ to zero. The best fit gave a $\chi^2= 224.6$ for 226 dof. The best-fit parameters are given in Table \ref{spectra_table}. Figure \ref{spectra_I_Q_U} shows the simultaneously fitted \textit{I} (top left panel), \textit{Q} (mid left panel) and \textit{U} (bottom left panel) spectra for all three DUs with their residuals in the adjacent right panels. Since {\sc{polpow}} has a power-law dependence of the polarization with energy, we integrated the model in the $2-8$ keV band and calculated the PD to be $4.23^{+3.59}_{-1.98} \%$ and PA to be $21.02^{\circ}\pm1.33^{\circ}$. The values estimated from the {\sc{PCUBE}} algorithm lie within the error bars of the ones estimated from the {\sc{xspec}} method, confirming that, compared to the HSS, the polarization of 4U 1630 decreased in the SPL state. 

\begin{table}
\centering
 \caption{The polarization degree (PD), and polarization angle (PA) for 4U~1630$-$47 extracted using the PCUBE algorithm.}
 \begin{center}
\scalebox{0.95}{%
\hspace{-0.5cm}
\begin{tabular}{ccccc}
\hline
 & DU1 & DU2 & DU3 & All DUs\\ \hline
PD (\%) & $6.7\pm{0.2}$ & $6.7\pm{0.4}$ & $7.1\pm{0.4}$ & $6.7\pm{0.2}$\\ 
PA (deg) & $20.8\pm{0.9}$ & $19.5\pm{1.6}$ & $21.3\pm{1.5}$ & $20.8\pm{0.9}$\\ 
Q/I (\%) & $5.0\pm{0.2}$ & $5.2\pm{0.4}$ & $5.3\pm{0.4}$ & $5.0\pm{0.2}$\\ 
U/I (\%) & $4.4\pm{0.2}$ & $4.2\pm{0.4}$ & $4.8\pm{0.4}$ & $4.4\pm{0.2}$\\ 
\hline
\end{tabular}}
\end{center}
\label{pol_table}
\end{table}

\begin{table}
\centering
 \caption{{\it{NICER}} and {\it{IXPE}} best-fitting spectral parameters for 4U~1630$-$47 with the model {\sc{polpow*tbabs*(thcomp*diskbb+gaussian+gaussian)}}.}
 \begin{center}
\scalebox{0.95}{%
\hspace{-1.5cm}
\begin{tabular}{lll}
\hline
 Component    & Parameter &  Value\\ \hline
polpow  & $A_{\rm norm}$ ($10^{-2}$) &  $1.7 \pm{0.3}$\\
polpow  & $A_{\rm index}$ &  $-0.6 \pm {0.1}$\\
polpow  & $\psi_{\rm norm}$ (degrees) & $21.0\pm{0.8}$\\
TBabs     & $N_H$ ($10^{22}$ cm$^{-2}$) &  $11.51\pm{0.02}$ \\
thcomp   & $\Gamma_\tau$  & $2.1\pm 0.2$\\ 
thcomp   & $cov_{frac}$ &  $0.5\pm 0.1$\\
diskbb  & $T_{in}$ (keV)  & $1.71 \pm {0.04}$\\
diskbb   & norm &  $223^{+21}_{-16}$ \\
gaussian & LineE (keV) & $7.07\pm 0.03$ \\
gaussian & Sigma (keV) ($10^{-2}$) &  $7.3^{+4.4}_{-6.9}$ \\
gaussian & flux ($10^{-3}$ ph cm$^{-2}$s$^{-1}$) &  $1.6\pm{0.5}$ \\
gaussian & LineE (keV) & $6.51\pm 0.12$ \\
gaussian & Sigma (keV)  & $0.15^{+0.17}_{-0.10}$ \\
gaussian & flux ($10^{-3}$ ph cm$^{-2}$s$^{-1}$) &  $1.2^{+0.9}_{-0.4}$ \\
Total Flux$^{\dagger}$ & ($10^{-8}$ erg cm$^{-2}$ s$^{-1}$) & $3.504\pm 0.003$ \\
\hline
$\chi^2$/dof & & 224.6 / 226 \\
\hline
\multicolumn{3}{l}{$^{\dagger}$ Total unabsorbed flux in the $2-12$ keV range.}
\end{tabular}}
\end{center}
\label{spectra_table}
\end{table}
\section{Discussion}\label{sec_concl}
In this work, we report for the first time a polarization fraction of $\sim$7~\% at a polarization angle of $\sim$21 $^\circ$ in the SPL of the blackhole transient 4U 1630. The $15-50$ keV {\it{MAXI}} lightcurve shows a significant rise in the hard X-ray flux, while the {\it {SWIFT}} and {\it {NICER}} lightcurves show an increase in the $2-20$ keV and $0.2-12.0$ keV band with respect to the previously reported HSS of the source \citep{ra23}.  In the {\it {NICER}} HID plot, the $0.2-12.0$ keV flux and the hardness ratio increase in the SPL state. The 2-12 keV {\it {NICER}} can be well fitted with a combination of a thermal and a comptonization component, the latter with spectral index $\Gamma \sim$2.1. The spectrum also shows a weak reflection feature at $\sim$7.0 keV. In the SPL state, we have observed an energy-dependent polarization fraction.\\

The polarization angle for 4U 1630 in the SPL state is consistent with the one reported for the HSS within less than 3 $\sigma$ (as shown in Figure \ref{2_8_stoke}). Also, the energy-dependent polarization fraction shows similar behaviour as observed for the HSS (see Figure \ref{ene_pol}) but is significantly lower than in the HSS. This shows that the corona is not the main cause of the polarimetric properties of 4U 1630. 
It should be noted that there is a significant overall drop of the polarization fraction from $\sim$8\% to $\sim$7\% in the SPL state compared to the HSS. This drop could be due to the partial depolarization by the corona in the SPL state. \citet{pou23} predicted that the polarization fraction could exceed 10\% for high inclination (i > 60°) BXBs, in the framework of an outflowing coronal model. However, the similarity of the energy-dependent polarization fraction in the SPL and HSS further eliminates any possibility of the corona being the source of polarization for 4U 1630.
Instead, in the SPL state, the polarization could arise as a combined effect of a slim disk and a partially ionized atmosphere moving with relativistic speed as proposed by \citet{Ratheesh_2023} for the HSS of 4U 1630.

The presence of a strong corona accompanied by a weak reflection feature argues against the origin of polarization as a combined effect of the accretion disk, a disk wind and a weak corona as reported by \citet{Kushwaha_2023}. 
Interestingly, for a high inclination BXB harbouring a Kerr black hole, the observed energy-dependent polarization fraction for the source could be explained with a thick disk model ($h/r=0.5$) (see Figure M 16 of \citealt{Ratheesh_2023}). But the thick disk model fails to explain the high polarization fraction reported in the present work for 4U 1630 as shown in Figure M 16 of \citealt{Ratheesh_2023}.

An interesting speculation would be that the polarization is due to self-irradiation of the disk around a Kerr black hole \citep{ca75,ca76}. Previous spectral studies reveal that 4U 1630 harbours a highly spinning black hole (\citealt{ki14}, \citealt{li22} and references within); in that case, the frame-dragging effect \citep{ba75} would be significant, leading to self-irradiation of the disk. Further, for 4U 1630, the similarity between the energy-dependent polarization fraction behaviour in the HSS and SPL states strengthens the explanation that the polarization is due to self-irradiation of the disk around a Kerr black hole \citep{ca75,ca76}.
Developing emission models for the direct and reflected components in black hole X-ray binaries is valuable in this context. Studying other X-ray binary sources in different states with future {\it {IXPE}} and {\it {XPoSat}} observations can provide insights into the polarization mechanism and the geometry of the disk and corona system.

\section*{Acknowledgements}
This research has used data from the High Energy Astrophysics Science Archive Research Center Online Service, provided by the NASA/Goddard Space Flight Center.
DR acknowledges Tomaso M. Belloni for providing the GHATS package used in this work for timing analysis.
MM acknowledges support from the research program Athena with project number 184.034.002, which is (partly) financed by the Dutch Research Council (NWO). 

\section*{Data Availability}
The {\it{NICER/XTI}} and \textit{IXPE} observations used in this work are available at the \href{https://heasarc.gsfc.nasa.gov/db-perl//W3Browse/w3browse.pl}{HEASARC website}.
\bibliographystyle{mnras}
\bibliography{manuscript} 
\label{lastpage}
\end{document}